\documentclass[11pt]{article}
\usepackage[margin=1in]{geometry}
\usepackage{graphicx,amsmath,booktabs,siunitx,microtype,xcolor}
\usepackage{caption}   
\usepackage[numbers,sort&compress]{natbib}
\usepackage[hidelinks]{hyperref}
\title{\textbf{Neuromorphic Decoding and Dynamical Image Reconstruction from Populations of Spiking Retinal Ganglion Cells}}
\author{Ning Zhang\thanks{ \href{mailto:ning_zhang1@brown.edu}{ning\_zhang1@brown.edu}}
  \ and Arto Nurmikko\thanks{\href{mailto:arto_nurmikko@brown.edu}{arto\_nurmikko@brown.edu}} \\
  \small School of Engineering, Brown University, Providence, RI 02912 USA}

\date{}

\begin{document}
\maketitle


\begin{abstract}
\noindent\textit{Objective.} Decoding visual stimuli from the spiking activity of retinal ganglion cells (RGCs) informs both our understanding of the retinal code and the design of visual prostheses. Natural vision is dynamic, yet state-of-the-art decoders mostly reconstruct single frames with artificial neural networks that process dense, continuous-valued signals rather than the retina's sparse, asynchronous spikes. We asked whether dynamic scenes can be reconstructed and tracked entirely within the spike domain, and at what computational cost.

\noindent\textit{Approach.} We built an end-to-end spiking neural network (SNN) whose neurons integrate the recorded RGC spikes over time. Populations of up to 1836 RGCs were recorded \emph{ex vivo} from mouse retina on high-density microelectrode arrays while flashed images, moving bars, drifting gratings and randomly moving handwritten digits were presented. For moving digits, one shared spiking encoder fed two readouts, one reconstructing the image and one decoding the digit's position. Compute energy was estimated from counted synaptic operations.

\noindent\textit{Main results.} The SNN decoded what and where at once. It reconstructed the stimuli with a structural similarity (SSIM) of up to 0.93, refining the reconstruction as spikes accumulated, and tracked a randomly moving target with a mean error of 1.33 pixels on a $32\times32$ grid. Its fidelity nearly matched that of the best of 17 larger spiking and artificial decoders (5.9 million to 1.7 billion parameters, versus 5.4 million), and at a mean firing rate of 14.2\% it needed an estimated $\sim$70-fold less compute energy than a dense artificial network of the same architecture.

\noindent\textit{Significance.} Dynamic visual scenes can be decoded from retinal population activity in the spike domain, with a compact network and at a compute cost compatible with implantable and wearable neural interfaces, supporting spike-native decoders for visual prostheses.
\end{abstract}

\section{Introduction}
The visual system is an existence proof that spiking, event-driven computation can transform incident light into perception at remarkable energy efficiency and low latency. In the vertebrate eye, the retina maps and compresses a photoreceptor image into parallel spike trains computed by a network of some $\sim$40 functionally distinct retinal ganglion cell (RGC) types~\citep{baden2016diversity,goetz2022unified}. Reconstructing the visual stimulus directly from recorded spiking activities of a large enough RGC subset is a dual scientific and engineering opportunity. First, a successful computational decoding model can improve our understanding of how the retina transforms and compresses visual information prior to transmission through the optic nerve to the brain. Second, a successful computational paradigm can help advance efforts to restore vision by guiding the stimulation strategies of next-generation retinal prostheses~\citep{wu2023prostheses,kim2021nonlinear}.

Two properties of biological vision make the RGC population decoding problem well suited to a direct neuromorphic computational approach. First, natural vision is overwhelmingly \emph{dynamic}: the retinal image changes continuously, with eye movements and with motion in the scene itself. A useful decoder must be able to generate a time-varying reconstruction of a time-varying scene rather than reproduce a single frozen snapshot. Second, the retinal code is fundamentally \emph{sparse and asynchronous}: information is carried by discrete spikes across a heterogeneous cellular population, not by dense, clocked, continuous-valued activations. A decoder that performs two simultaneous functions on retinal population data (reconstructing a visual target and tracking its motion) while running entirely in the spike domain represents, we suggest, a more direct reflection of the underlying biology, and it offers potential savings in computational energy.

Currently, the dominant paradigm for neural decoding of RGC population signals is an artificial neural network (ANN) model operating on synchronous, continuous-valued signals. Linear, Bayesian and nonlinear decoders have shown that natural images can be recovered from RGC populations~\citep{parthasarathy2017bayesian,brackbill2020reconstruction,kim2021nonlinear}, and deep networks, from the Spike-Image Decoder to diffusion-prior systems developed for functional imaging, have pushed reconstruction quality substantially higher~\citep{zhang2020sid,takagi2023ldm,scotti2024mindeye2,chen2023mindvideo,yu2024dynamic}. These decoders can indeed achieve high fidelity, but in our view they are not naturally matched to the biological substrate in ways that matter for energy-efficient neural engineering. First, such decoders are largely \emph{energy-blind}: they are trained, and their outcomes usually reported, without regard to the computing cost, which matters in any implantable or wearable interface~\citep{horowitz2014energy,rathi2022neuromorphic,evenchen2020power}. Second, they are \emph{representationally mismatched}: the ANNs process dense analog tensors rather than the sparse asynchronous spikes that carry information in the nervous system. This mismatch can limit both biological interpretability and downstream deployment on future event-driven neuromorphic hardware~\citep{davies2018loihi,yao2024speck}. Third, and perhaps most consequential in the inherently nonstationary moving visual world, they have largely addressed only \emph{single, non-moving} images. Even powerful \emph{linear} decoders, which reconstruct the stimulus as a weighted sum of the population’s spike counts using one fixed spatial filter per cell, recover images from primate RGC data that are “more similar to the smoothed image than to the original image”~\citep{brackbill2020reconstruction}: fine detail is already lost before the target begins to move. Once the target moves, the degradation compounds: with one fixed spatial filter per cell and no mechanism to integrate evidence over time, the recovered image is further blurred by the target's own motion. The dimension that matters most for natural vision is thus the one such decoders handle least well.

In this paper we address these issues by introducing a novel, adaptable, biomimetic decoder built on a spiking neural network (SNN). The model is applied to multielectrode recordings from ex vivo mouse retina with data from up to 1836 spiking RGCs (ON, OFF and ON--OFF cells). Our model is capable of dynamical image reconstruction and simultaneous target tracking; it refines the reconstruction through its own temporal dynamics, and it does so efficiently.

By way of background, directly trained deep SNNs have recently become computationally practical through surrogate-gradient learning~\citep{neftci2019surrogate,eshraghian2023training,fang2023spikingjelly} and spiking transformers now approach ANN accuracy~\citep{zhou2023spikformer}, with spike-driven variants retaining sparse, addition-only computation~\citep{yao2023spikedriven,yao2024metaspikeformer}. Work on spiking image reconstruction and tracking, however, has centered primarily on robotic vision applications using data from event-driven optical cameras~\citep{zhu2022evsnn,ji2023sctn}, where computing efficiency is typically assessed post hoc. We have recently applied an SNN model to reconstruct and track moving targets from event-camera streams when imaging through optically dense scattering media~\citep{zhang2026}. By contrast, the substrate in this paper is the biological retina. Its RGC spiking populations, recorded by dense microelectrode arrays, are receptive-field mapped and separated by contrast polarity, and they carry the trial-to-trial variability inherent to living tissue, as opposed to the deterministic protocol of man-made cameras.

\begin{figure}[htbp]\centering
  \includegraphics[width=\linewidth]{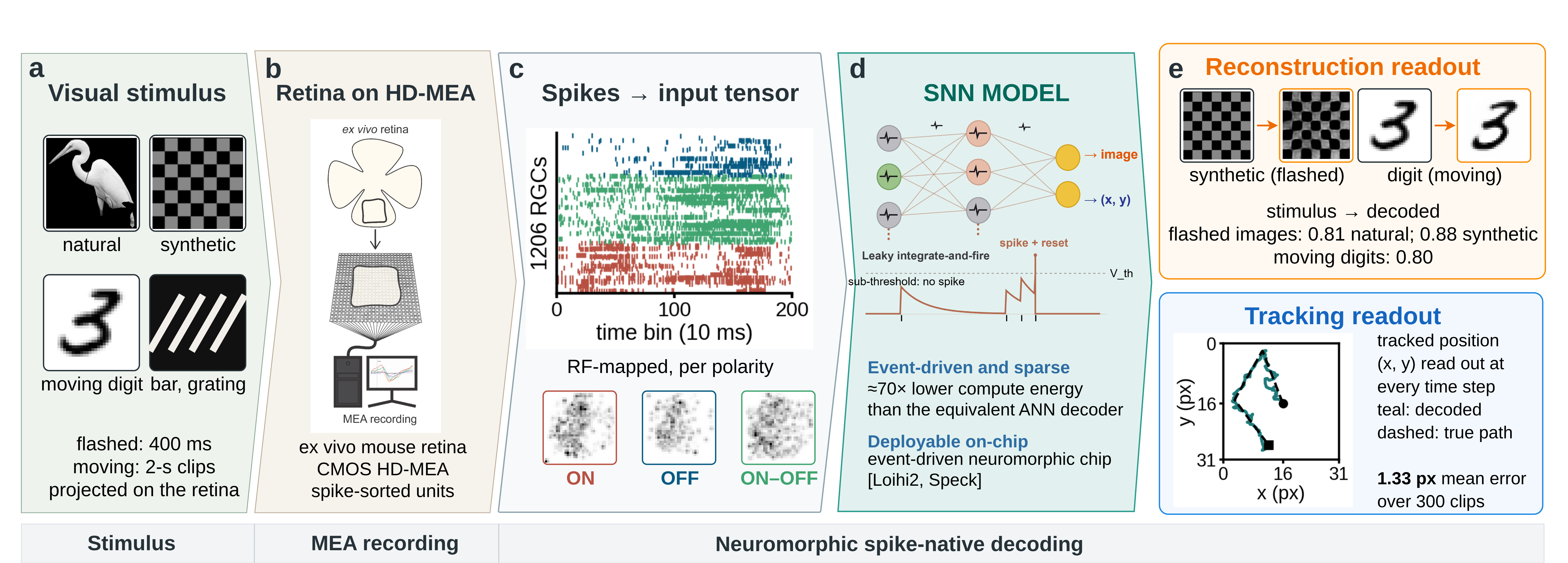}
  \caption{End-to-end neuromorphic decoding of dynamic vision from retinal ganglion cell populations. (a) Visual stimuli, flashed natural or synthetic images (400~ms) or moving targets (2-s clips of randomly moving digits, bars and gratings), are projected onto (b) an ex vivo mouse retina on a CMOS high-density microelectrode array, yielding spike-sorted RGC units. (c) Population spikes (raster of one held-out clip; colors mark ON, OFF and ON--OFF cells) are placed at each cell's receptive-field center with a Gaussian kernel, giving a $[T\times3\times32\times32]$ input tensor with one channel per polarity. (d) A single spike-native decoder, a shared spiking encoder with a reconstruction readout and a tracking readout, built from leaky or non-leaky integrate-and-fire (LIF/IF) neurons and trained with surrogate gradients; its sparse, event-driven computation uses a small fraction of the compute energy of the equivalent ANN decoder (Sec.~\ref{sec:energy}) and can map onto neuromorphic hardware. (e) Its outputs: the reconstructed stimulus (a flashed synthetic pattern and a moving digit; masked-SSIM 0.81 natural and 0.88 synthetic, SSIM 0.80 for moving digits) and the target position at every time step (decoded, teal; true path, dashed; 1.33~px mean error over 300 held-out clips). The moving-digit example is the same held-out clip as in Figs.~\ref{fig:arch} and~\ref{fig:dynamic}.}
  \label{fig:concept}
\end{figure}

The end-to-end spiking decoder introduced in this paper is a compact spiking network algorithm capable of reconstructing images of a dynamical target while tracking its position with the same architecture. Its computational strategy consumes a fraction of the energy of an equivalent ANN model, a direct reflection, we believe, of the benefit of the SNN model’s mimicking the computational principles of the retina itself. The overall flow of the ‘spike-native’ SNN approach is shown schematically in Fig.~\ref{fig:concept}. We show below (1) how the approach has allowed us to demonstrate what is, to our knowledge, the first end-to-end spiking neural population decoder that reconstructs \emph{dynamic, moving} visual targets, as well as flashed images, from real RGC population spikes. In a way that seems similar to how the brain decodes and interprets retinal data, the reconstruction is seen to sharpen in real time through temporal integration in the \emph{spike-native}, high-fidelity, compact network model. We also show (2) how the SNN engine, as a single shared spiking encoder, tracks the moving target’s position using a learned spiking readout that outperforms a calibrated population-vector (center-of-mass) readout~\citep{georgopoulos1986population} roughly two-fold, showing that spiking-based computation extracts position information that a classical readout leaves behind; (3) that the SNN engine, being spike-native, is energy efficient ($\sim$70$\times$ lower compute energy than the ANN-equivalent); and (4) we suggest that the model can be compatible with emerging neuromorphic hardware~\citep{davies2018loihi,yao2024speck}.

While this work focuses on decoding retinal cellular populations, the approach can potentially extend to the large-scale decoding of neural data required by brain-computer interfaces that deploy dense microelectrode arrays and probes~\citep{jun2017neuropixels,musk2019neuralink,obaid2020microwire,jung2025wireless} or distributed microchip sensors~\citep{lee2024asynchronous}. The model’s energy efficiency suggests that it could scale to decoding cortical dynamics from thousands of neurons.

\section{Results}
\begin{figure}[htbp]\centering
    \includegraphics[width=\linewidth]{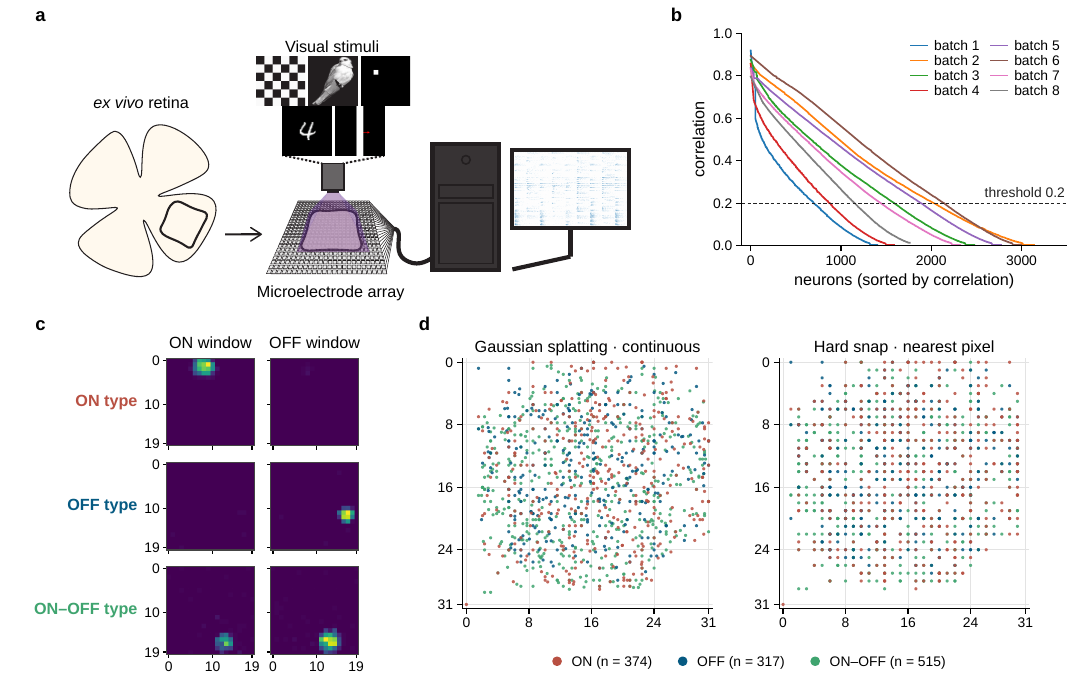}
    \caption{\textit{Ex vivo} retina MEA recording and data preparation. (a) Schematic of \textit{ex vivo} retina MEA recording testbed. Up to 1836 cells were recorded from a piece of retina (2~mm $\times$ 2~mm) by the array, with visual stimuli delivered by a Digital Light Processor (DLP) at a wavelength of 405~nm. (b) After spike sorting, the mean pairwise correlation coefficient across trials was calculated for each unit, and units with a mean correlation coefficient greater than 0.2 (dashed line) were retained as high-quality units; one curve per recording batch, units ranked by correlation. (c) Receptive fields (RF) of three representative cells of different types: responses over the $20\times20$ mapping grid in the ON and OFF windows. (d) Receptive-field centers of the 1206 cells of the moving-digit recording on the $32\times32$ input grid, colored by response polarity (ON, OFF, ON--OFF; the color key of (c) and Figs.~\ref{fig:concept}c and~\ref{fig:dynamic}a). Left: the continuous (sub-pixel) centers, at which each cell's firing rate is rendered with a unit-mass Gaussian kernel (Gaussian splatting; Methods) to form the decoder input. Right: the same centers snapped to the nearest grid pixel (hard snap), shown for comparison; cells that share a pixel are merged.}
    \label{fig:mea}
\end{figure}

All results are obtained with the spiking neural network model applied to spike trains of mouse RGCs recorded by a microelectrode array (MEA) (Fig.~\ref{fig:mea}a and Methods). High-quality single units were identified through spike sorting and quality control (pairwise correlations in Fig.~\ref{fig:mea}b), and each unit was annotated with receptive-field location information for the ON, OFF and ON--OFF principal cell types (Fig.~\ref{fig:mea}c). We map per-cell firing rates onto a $32\times32$ grid at each cell's receptive-field (RF) center using Gaussian kernels (Fig.~\ref{fig:mea}d), and separate cells into ON, OFF and ON--OFF channels, giving an input tensor of shape $[\,T\times C\times 32\times 32\,]$ (Methods).

\begin{figure}[htbp]\centering
  \includegraphics[width=0.95\linewidth]{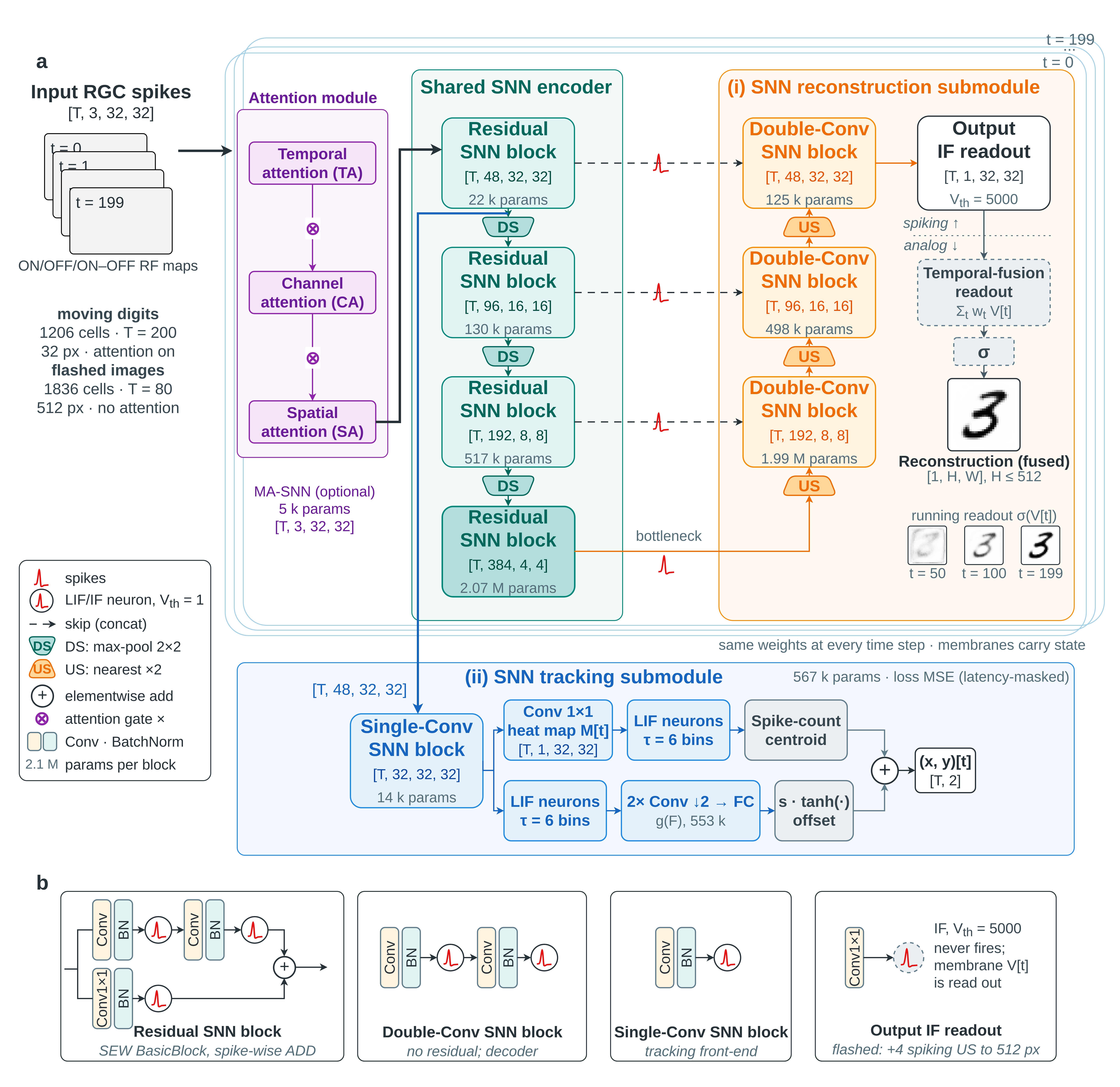}
  \caption{Shared-encoder dual-head spiking network architecture. (a) Input spike maps (1206 RGCs, ON/OFF/ON--OFF channels, $T=200$ bins) pass through an optional MA-SNN attention front-end~\citep{yao2023masnn} (purple; moving-digit models only) into a shared SEW-ResNet~\citep{fang2021sew} spiking U-Net encoder (teal; DS, $2\times2$ max-pool). Its multiscale features feed (i) the reconstruction submodule (orange; US, nearest-neighbor $\times2$ with concatenated skips), whose output neuron never fires and whose membrane $V[t]$ is fused over time and passed through a sigmoid, and (ii) the tracking submodule (blue), which adds a full-resolution regression offset to a spike-count centroid. Apart from the input stage (the optional attention module and the first convolution, which read the RF-mapped input), every stage is spiking (LIF/IF neurons; red symbols mark spike tensors); gray blocks are analog readout operations and the dotted line marks the spiking/analog boundary. The same weights are applied at every time step (nested frames). Parameter counts are printed per block: 5.9~M for the moving-digit model and 5.4~M for the flashed-image model (1836 cells, $T=80$, no attention, 512-pixel output). Thumbnails: running readout $\sigma(V[t])$ at $t=50$, 100 and 199, and the fused reconstruction of one held-out clip (digit 3). (b) Block definitions and legend.}
  \label{fig:arch}
\end{figure}

The SNN decoder module is based on a U-Net-like architecture~\citep{ronneberger2015unet} (Fig.~\ref{fig:arch}) built from leaky or non-leaky integrate-and-fire (LIF/IF) neurons. Its encoder processes the input along its own native time axis, so that evidence about the stimulus accumulates across time bins rather than being pooled in a preprocessing step. Two modules read the same shared features: an SNN reconstruction module that outputs the image, and an SNN tracking module that outputs the target's $(x,y)$ position. We evaluated two recording regimes: (a) recordings from 1836 spiking RGCs responding to 4000 flashed natural and synthetic images and (b) recordings from 1206 RGCs responding to 3000 clips of randomly moving handwritten digits, moving bars, and drifting gratings (bars and gratings from a separate recording of 760 RGCs). All reported numbers are computed on held-out data (Methods).

\subsection{Image reconstruction -- flashed stimulus}\label{sec:flashed}
\begin{figure}[htbp]\centering
  \includegraphics[width=\linewidth]{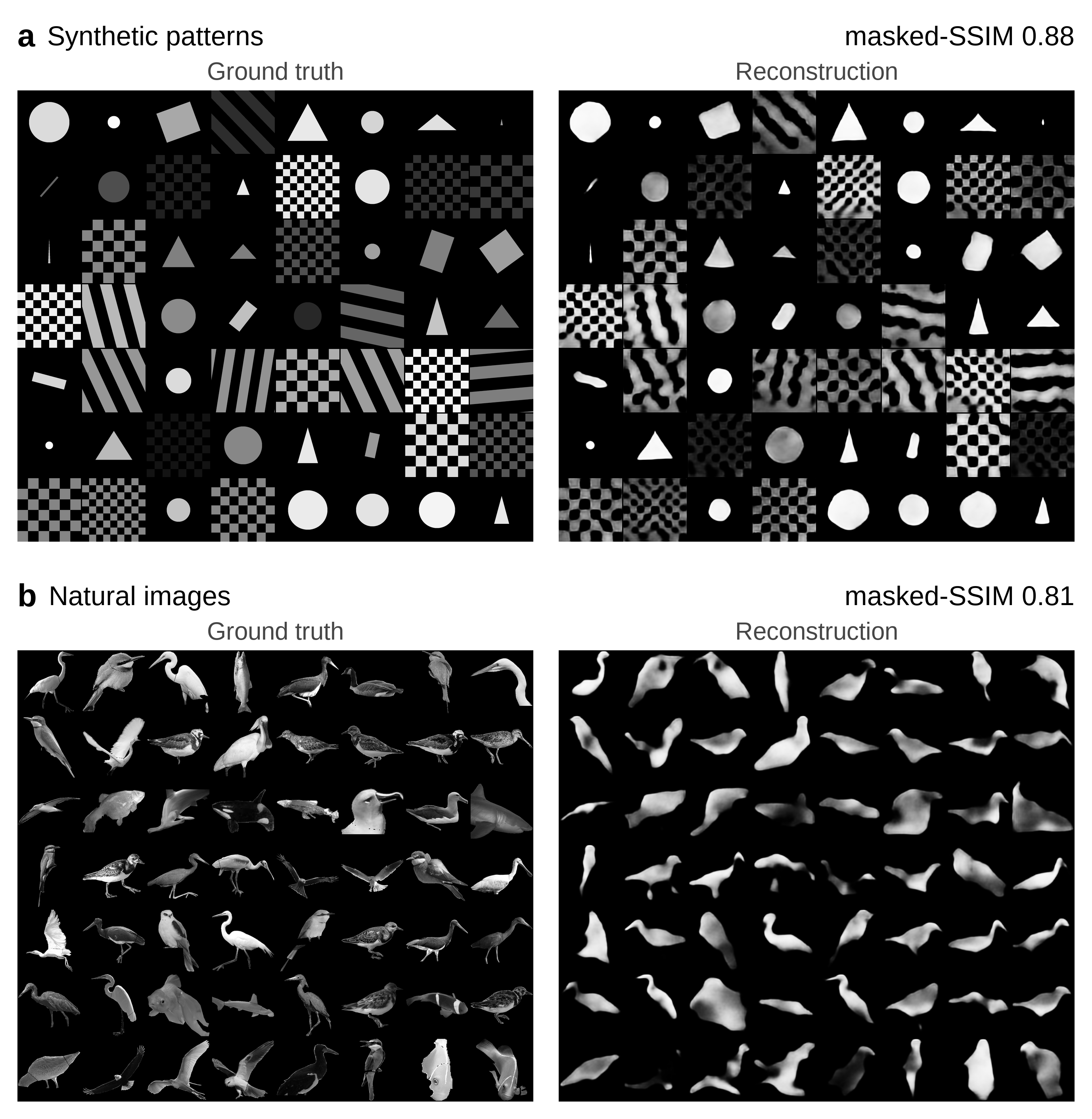}
  \caption{Image reconstruction of flashed stimuli decoded from 1836 RGCs by the end-to-end spiking U-Net. Ground truth (left) versus spike-native reconstruction (right), $7\times8$ held-out examples per panel, for (a) synthetic patterns (masked-SSIM 0.88) and (b) natural images (masked-SSIM 0.81). Similarity index is quoted over all 200 held-out images.}
  \label{fig:flashed}
\end{figure}

The end-to-end SNN engine was applied to reconstruct flashed stimuli (each image projected as a 400~ms flash followed by 400~ms of darkness) from the full recorded population of 1836 RGCs. The reconstructions reach reasonably high fidelity in structural similarity: masked-SSIM $\sim$0.81 on natural images and $\sim$0.88 on synthetic patterns (Fig.~\ref{fig:flashed}). The image is not decoded in a single step: the reconstruction builds up as spikes accumulate over the trial and keeps sharpening after the image is switched off, as the offset responses arrive (Video~S3).

\subsection{Reconstruction fidelity versus model size}\label{sec:sota}
\begin{figure}[htbp]\centering
  \includegraphics[width=\linewidth]{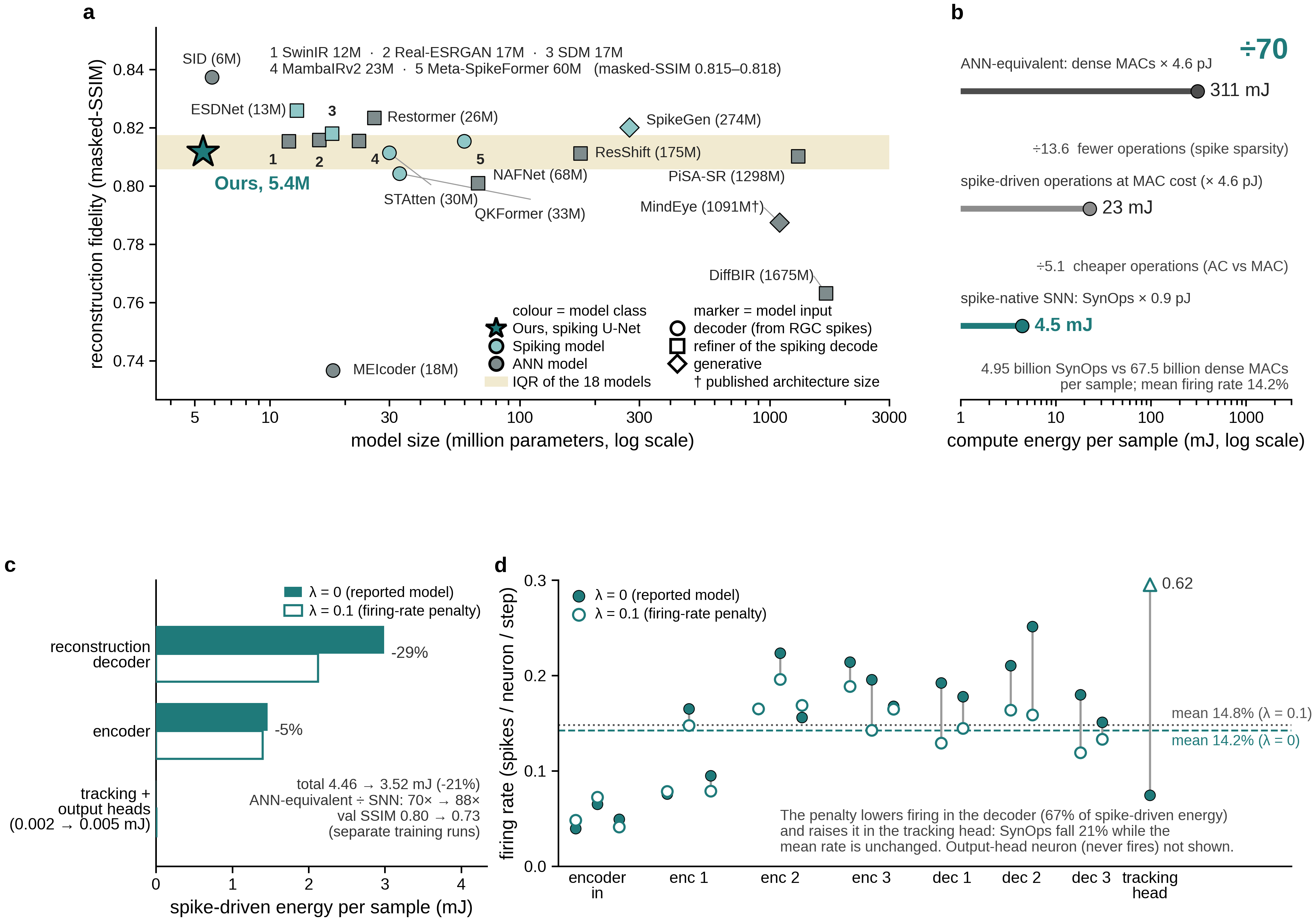}
  \caption{Reconstruction fidelity versus model size, and compute energy. (a) Reconstruction fidelity (masked-SSIM, window 11, 512~px, $n=200$ held-out images, identical protocol for every point) versus model size for 18 decoders on the same 1836-RGC flashed natural-image task: our spiking U-Net (teal star), other spiking models (light teal) and artificial networks (gray); circles decode the RGC input directly, squares refine the Meta-SpikeFormer decode, diamonds are generative models. Fidelity saturates in a narrow band (shaded, interquartile range of the 18 models) across more than two orders of magnitude in parameter count, suggesting a neural-data-limited rather than capacity-limited regime; the spiking U-Net of this paper (star, $\sim$5.4M parameters) reaches this band with $5.5$--$11\times$ fewer parameters than the spiking-transformer decoders. Models 1--5 lie within 0.003 SSIM of each other and are numbered (markers 2 and 3, both 17M, are offset horizontally for visibility); $\dagger$ published architecture size; spiking-model parameters measured by torchinfo. (b) Compute energy per sample under a 45~nm convention (multiply--accumulate 4.6~pJ, accumulate 0.9~pJ): the ANN-equivalent (dense multiply--accumulates), the same spike-driven operation count at multiply--accumulate cost, and the spike-native network (synaptic operations at accumulate cost); the $\sim$70$\times$ reduction factorizes into $\sim$13.6$\times$ fewer operations (spike sparsity at 14.2\% mean firing) and $\sim$5.1$\times$ cheaper operations. (c) Spike-driven energy by module for the reported model ($\lambda=0$) and a model trained with a firing-rate penalty ($\lambda=0.1$): the penalty removes 21\% of the synaptic operations ($\sim$70$\times$ $\to$ $\sim$88$\times$) at a reconstruction cost (validation SSIM 0.80 $\to$ 0.73; separate training runs). (d) Firing rate of every spiking layer for the two models: the penalty lowers firing in the decoder, where fan-out is largest, and raises it in the tracking head, so the mean rate is unchanged while synaptic operations fall. The output-head neuron never fires and is not shown. Energy estimates are compute-only (memory movement excluded).}
  \label{fig:efficiency}
\end{figure}

To place the image reconstruction results in context, we benchmarked a suite of state-of-the-art reconstruction architectures applied to the \emph{same} held-out 1836-cell natural-image decoding task: spiking decoders (Meta-SpikeFormer~\citep{yao2024metaspikeformer}, QKFormer~\citep{zhou2024qkformer}, and STAtten, a spatio-temporal-attention spiking transformer~\citep{lee2025statten}) and state-of-the-art artificial restoration and decoder networks (SwinIR~\citep{liang2021swinir}, Restormer~\citep{zamir2022restormer}, Real-ESRGAN~\citep{wang2021realesrgan}, NAFNet~\citep{chen2022nafnet}, and the Spike-Image Decoder family~\citep{zhang2020sid}), together with ten further spiking and artificial decoders, refiners and generative models (Fig.~\ref{fig:efficiency}a). Each decoder was fed the identical receptive-field-mapped RGC input and scored with masked-SSIM. Table~\ref{tab:sota} lists performance against model size for our model, the three spiking decoders and four of the artificial networks, and Fig.~\ref{fig:efficiency}a plots it for all models. Two observations can be made. First, reconstruction fidelity appears to “saturate”: across 18 ANN and SNN architectures spanning more than two orders of magnitude in size, the masked-SSIM values fall in a narrow range, suggesting that the fidelity is limited by the information conveyed in the neural data rather than by a given model’s capacity. This is consistent with the flat spread we observe across attention variants, and with the loss of fine detail reported for linear decoding of primate RGC populations~\citep{brackbill2020reconstruction}. Second, and of practical consequence, the differentiating figure of merit is not only fidelity but \emph{efficiency}: our spiking decoder reaches the high-fidelity range with only $\sim$5.4M parameters, roughly $5.5$--$11\times$ fewer than the spiking-transformer decoders (QKFormer 33.0M, Meta-SpikeFormer 59.8M) and below the size of the artificial restoration networks. Reaching comparable fidelity at a fraction of the parameter count, in the spike domain, is the key result of this comparison: ours is a compact, energy-efficient SNN for the neural decoding task at hand.
 
\begin{table}[htbp]\centering\small
\caption{Reconstruction fidelity versus model size on flashed natural-image decoding from 1836 RGC spiking neurons. All models decode the same held-out receptive-field-mapped input. Parameters are measured by torchinfo for the spiking decoders; $\dagger$ denotes published architecture size for the artificial-network refiners. Masked-SSIM values are given on the shared held-out test set.}\label{tab:sota}
\begin{tabular}{llcc}
\toprule
Model & Domain & Params (M) & Masked-SSIM \\
\midrule
\textbf{Spiking U-Net (ours)} & spiking & \textbf{5.4} & \textbf{0.81} \\
QKFormer~\citep{zhou2024qkformer} & spiking & 33.0 & 0.80 \\
STAtten~\citep{lee2025statten} & spiking & $\sim$30 & 0.81 \\
Meta-SpikeFormer~\citep{yao2024metaspikeformer} & spiking & 59.8 & 0.82 \\
\midrule
SwinIR$^\dagger$ & ANN & 11.9 & 0.82 \\
Real-ESRGAN$^\dagger$ & ANN & 16.7 & 0.82 \\
Restormer$^\dagger$ & ANN & 26.1 & 0.82 \\
NAFNet$^\dagger$ & ANN & 67.9 & 0.80 \\
\bottomrule
\end{tabular}
\end{table}

\subsection{Image reconstruction -- moving objects, simple geometry}
\begin{figure}[htbp]\centering
  \includegraphics[width=0.96\linewidth]{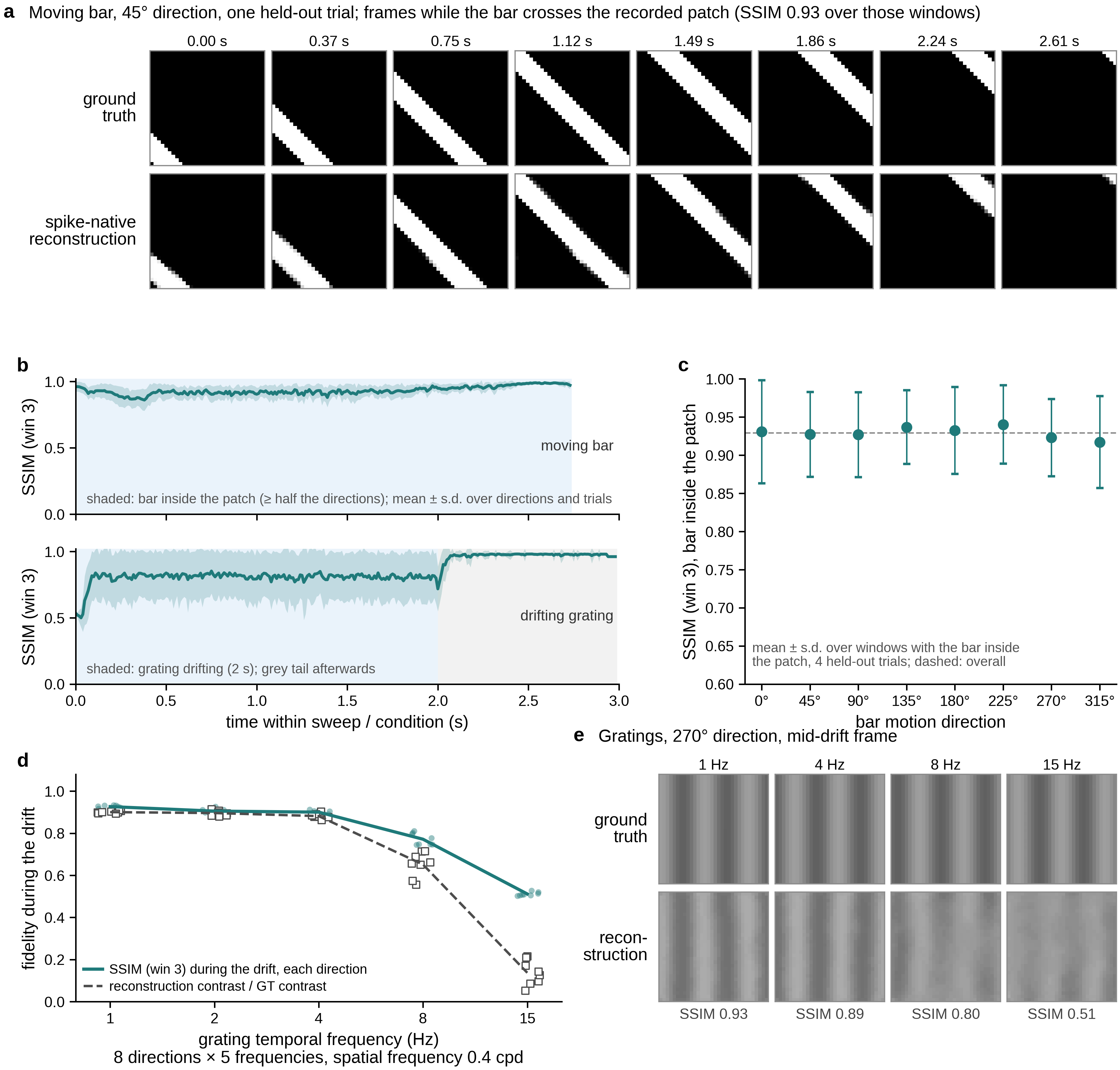}
  \caption{Reconstruction of moving bars and drifting gratings recorded and decoded from spiking RGC populations. (a) Moving bar, 45$^\circ$ (diagonal) direction, one held-out trial: ground truth (top) and spike-native reconstruction (bottom) at eight times while the bar crosses the recorded patch. (b) SSIM (window 3) versus time within a sweep (bars, top) or condition (gratings, bottom), mean $\pm$ s.d.\ over directions or conditions and trials; shading marks the bar inside the patch and the 2~s drift (light blue) and the blank after the drift (gray). (c) Bars: SSIM per motion direction over windows with the bar inside the patch (mean $\pm$ s.d., four held-out trials). (d) Gratings: SSIM during the drift (teal) and reconstruction-to-stimulus contrast ratio (open squares) for all 40 conditions (8 directions $\times$ 5 temporal frequencies at a spatial frequency of 0.4 cycles per degree); lines join means over directions. Reconstruction is faithful up to 4~Hz, degrades at 8~Hz and fails at 15~Hz. (e) Ground truth (top) and reconstruction (bottom) mid-drift for one direction at 1, 4, 8 and 15~Hz, with the SSIM of each condition. The reconstructions are the model output at the end of each causal sliding window (10 bins, 100~ms, for the bars; 3 bins, 30~ms, for the gratings; Methods).}
  \label{fig:geometry}
\end{figure}

As the first test of our SNN model on dynamical scenes, we asked the spike-integrating encoder to reconstruct simple classic patterns of moving geometry~\citep{hubel1959receptive,enrothcugell1966contrast}. The model achieved high fidelity: SSIM 0.93 for moving bars (four held-out trials, windows with the bar inside the recorded patch) and 0.80 for drifting gratings during the drift (0.86 including the blank interval after each drift; Fig.~\ref{fig:geometry}). Fidelity is independent of motion direction but falls with the gratings' temporal frequency, from 0.93 at 1~Hz to 0.51 at 15~Hz (Fig.~\ref{fig:geometry}d, Video~S1). The reconstruction quality for these simple patterns, which move along predictable, deterministic linear paths, showed that the encoder's temporal integration was able to follow a continuously changing visual scene, extending the results for the spatially static, temporally flashed images of Sec.~\ref{sec:flashed}.

\subsection{Image reconstruction -- randomly moving targets}
\begin{figure}[htbp]\centering
  \includegraphics[width=\linewidth]{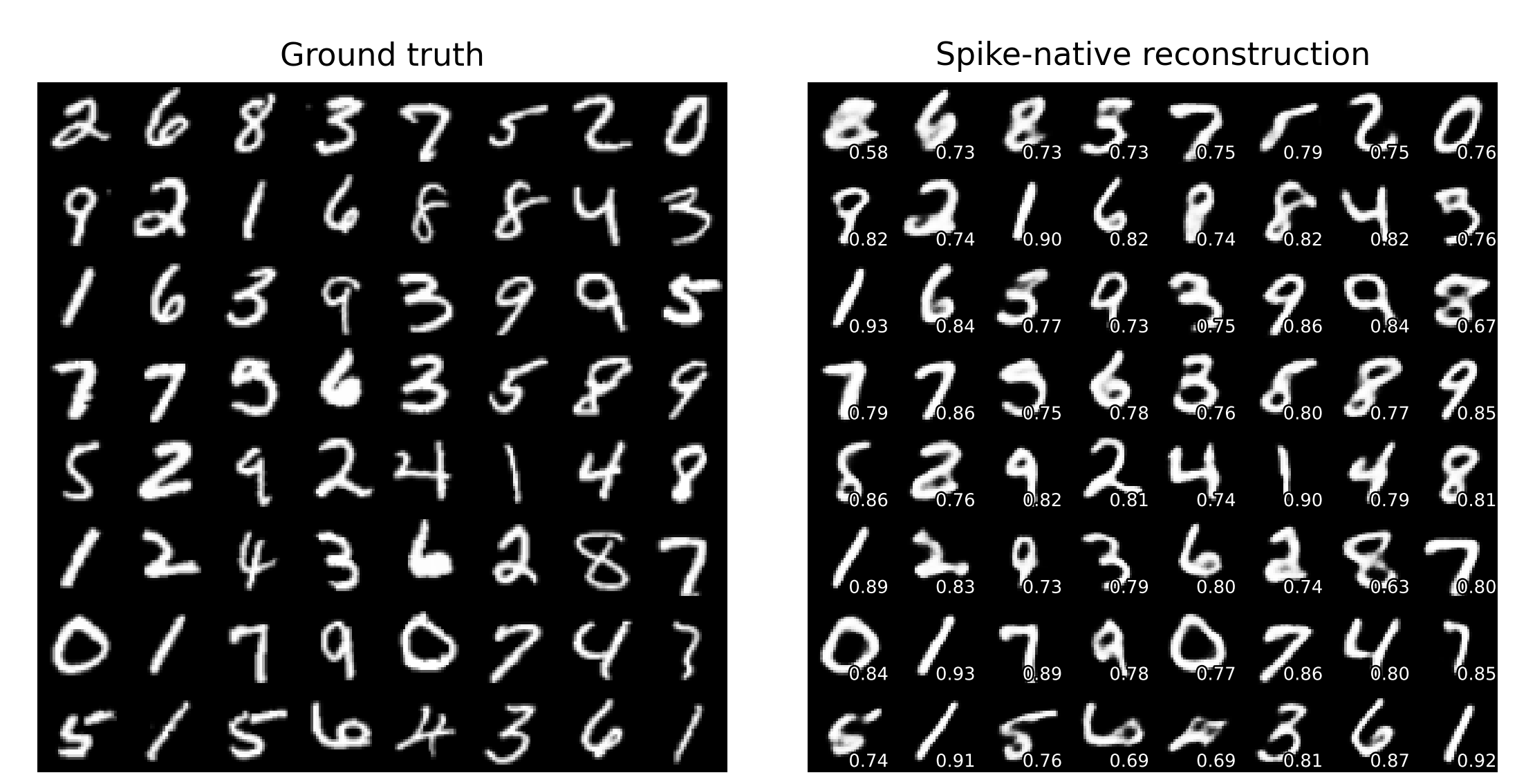}
  \caption{Consistency of dynamic reconstruction across the test set. Ground truth (left) versus spike-native reconstruction (right) for 64 randomly selected held-out clips of randomly moving digits, each following a distinct random $x$--$y$ trajectory, displayed as an $8\times8$ panel. Reconstruction is the temporally fused readout at the end of each clip. The SSIM (window-3) of each reconstruction is printed on its tile; over all 300 held-out clips the SSIM is 0.80~$\pm$~0.07 (mean~$\pm$~s.d.).}
  \label{fig:gallery}
\end{figure}

We then applied the model to the harder case of randomly moving visual targets, here handwritten MNIST digits~\citep{lecun1998mnist}. The objective was to reconstruct a target that moves unpredictably across the visual field, a regime that single-frame decoders cannot address. With video clips of randomly moving handwritten digits projected onto the ex vivo mouse retina, our SNN model reached a reconstruction fidelity of $\sim$0.80 SSIM, and the same shared-encoder model also tracked the target. The reconstructions are consistent across the held-out test set (Fig.~\ref{fig:gallery}), as summarized per digit class in Fig.~\ref{fig:dynamic}e. This is the main result of this paper. The reconstruction is \emph{refined over time by the network's own dynamics}: at early timesteps the estimate is diffuse, and the moving digit emerges as the membrane potentials of the spiking network accumulate evidence \emph{during} motion, with no explicit re-centering of the target (Fig.~\ref{fig:dynamic}a--c, Video~S2). The image sharpening is \emph{systematic} across the held-out set: averaged over all 300 test clips, reconstruction SSIM rises from near zero in the first 100~ms to $\sim$0.80 by the end of the 1.6-s motion phase, and then holds through the subsequent blank (Fig.~\ref{fig:dynamic}d). The sharpening thus follows the accumulation of spiking evidence rather than a single informative frame. We attribute the effect to the SNN's \emph{temporal memory integration}: the spiking membranes accumulate incoming evidence as potential over the video clip. On this polarity-labeled recording, a cell-type analysis shows that ON cells dominate decoding and that separating cells by response polarity contributes more to reconstruction quality than raw cell count, consistent with the parallel ON/OFF organization of retinal output~\citep{wassle2004parallel,krieger2017alpha}. 

\begin{figure}[p]\centering
\includegraphics[width=0.97\linewidth]{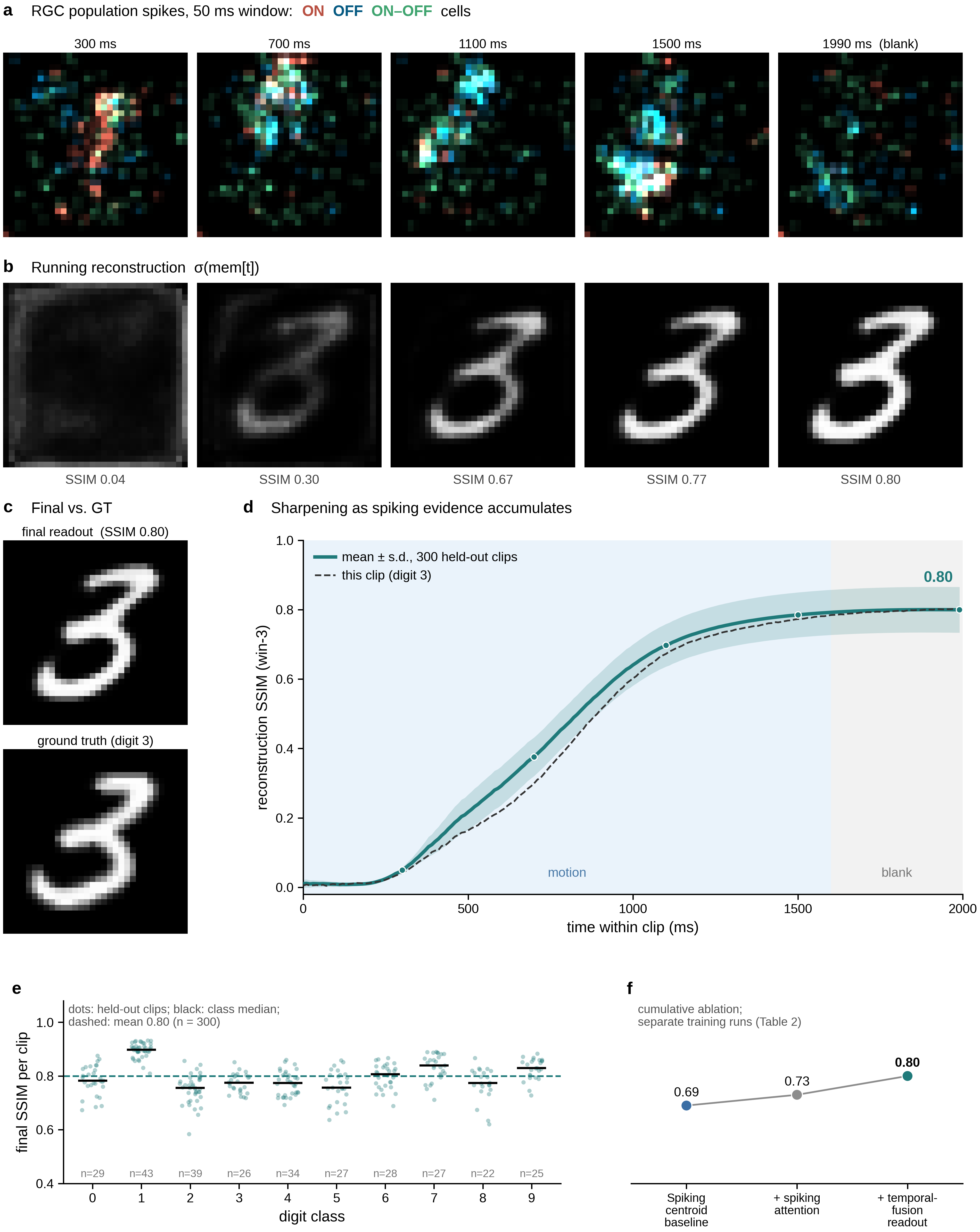}
\caption{Dynamical reconstruction of the image of a single, randomly moving digit projected on mouse retina, acquired from the shared spiking encoder (held-out clip, here for digit “3”), and the sharpening of the reconstruction over the whole held-out set. Panel descriptions continue on the following page.}
\label{fig:dynamic}
\end{figure}
\begin{figure}[t]
\ContinuedFloat
\caption{(Continued.) (a) RGC population spike maps over time (ON, OFF and ON--OFF cells in terracotta, blue and green, the color key of Figs.~\ref{fig:concept}c and~\ref{fig:mea}; spikes accumulated over 50~ms windows). (b) The running reconstruction refines over time: the image is diffuse and of low contrast early into the video clip; the moving digit emerges as the spiking membranes accumulate evidence during motion (SSIM against the ground-truth digit under each frame). (c) The final reconstruction closely matches the ground-truth digit. (d) Reconstruction sharpens systematically as spiking evidence accumulates: mean $\pm$ s.d.\ reconstruction SSIM (window-3, versus the ground-truth digit) of the running membrane readout $\sigma(\mathrm{mem}[t])$ at each timestep, averaged over all 300 held-out clips (teal), together with the curve of the clip shown in (a)--(c) (dashed). The estimate is diffuse early in the clip (SSIM near zero at 100~ms), rises through the 1.6~s motion phase as the spiking membranes integrate evidence, and plateaus at $\sim$0.80 by the end of motion, holding through the 0.4~s blank (shaded). The plateau matches the reported single-forward-pass reconstruction quality, consistent with the sharpening reflecting temporal evidence accumulation rather than a single informative frame. (e) Final SSIM of every held-out clip by digit class (dots), the median per class (black) and the overall mean (dashed; $n=300$). (f) Cumulative reconstruction ablation (Table~\ref{tab:abl}; separate training runs): spiking centroid baseline, adding the attention front-end, adding the temporal-fusion readout. Position tracking from the same encoder is quantified and assessed separately in Fig.~\ref{fig:track}.}
\end{figure}

\subsection{Object tracking}\label{sec:tracking}
\begin{figure}[htbp]\centering
  \includegraphics[width=\linewidth]{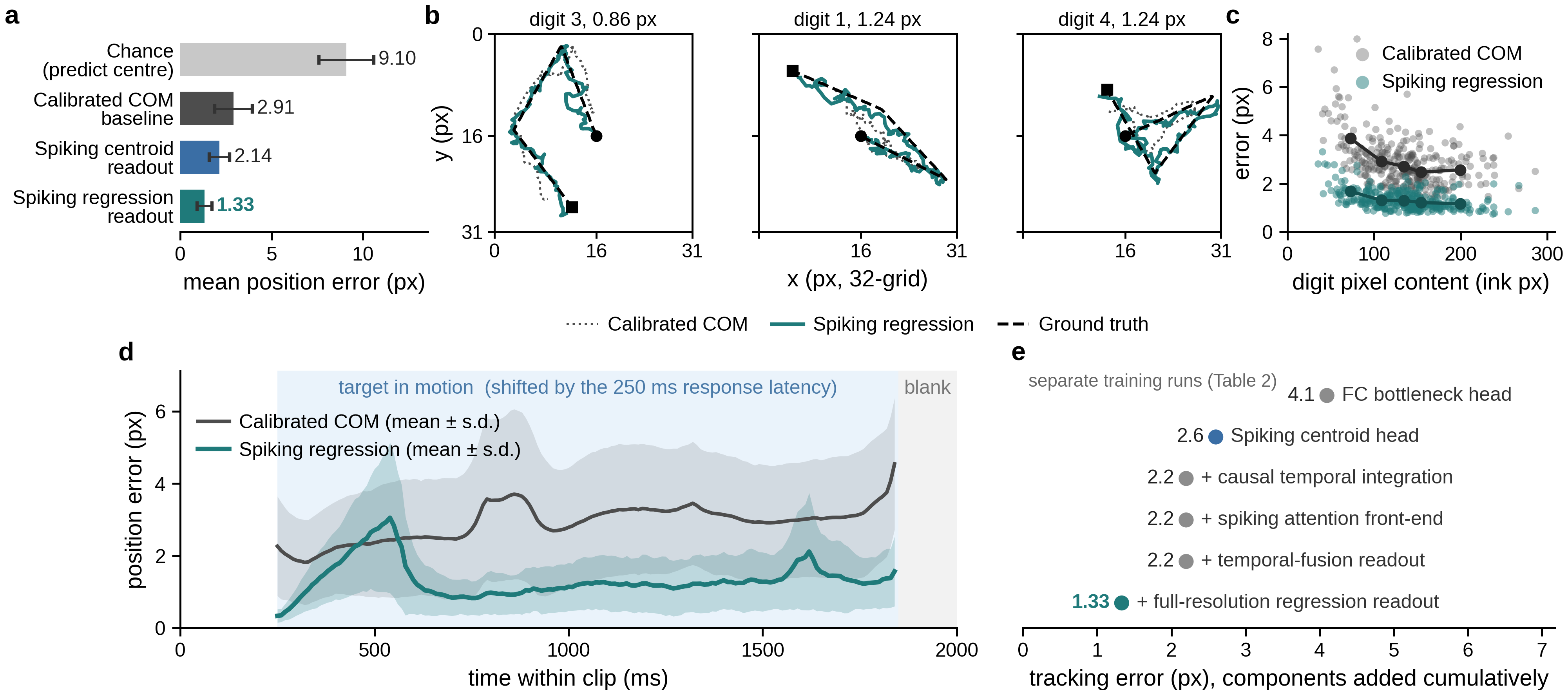}
  \caption{The SNN outperforms a calibrated center-of-mass readout in tracking a randomly moving object. (a) Mean $\pm$ s.d.\ position error (32-grid px) over the 300 held-out clips for each readout: chance (predict-center, 9.10~px), the calibrated center-of-mass (COM) baseline (2.91~px), the spike-count centroid readout (2.14~px) and the full-resolution spiking regression readout (\textbf{1.33~px}), which beats the calibrated COM baseline roughly two-fold. (b) True trajectory (dashed black; circle, start; square, end) overlaid with the decoded trajectory of the spiking regression readout (teal) and of the calibrated COM (dotted gray) for three held-out clips; the error of the spiking readout is printed above each. (c) Per-clip error versus the digit's pixel content (ink pixels) for the calibrated COM (gray) and the spiking regression readout (teal); lines join the quintile means. Error falls as pixel content grows for both readouts, and the spiking readout stays below the COM baseline at every size. (d) Position error versus time within the clip, mean $\pm$ s.d.\ over clips, for the spiking regression readout and the calibrated COM; the shaded window is the target's motion shifted by the 250~ms response latency, and curves are shown where a target position is defined. (e) Cumulative component ablation of the tracking readout (Table~\ref{tab:abl}; separate training runs).}
  \label{fig:track}
\end{figure}

The SNN tracking module, which recovers the target's position, is embedded in the same shared encoder that reconstructs images of randomly moving targets (here handwritten digits). We asked how the model performs with reduced, sparse information. Figure~\ref{fig:track}c plots the performance of the tracking module against different levels of pixel content. A simple spike-count centroid readout reaches a reasonable mean position error of 2.14 pixels (px). As an alternative, a full-resolution spiking regression readout using the entire feature map lowers the error to \textbf{1.33~px} on held-out clips (Fig.~\ref{fig:track}a,b), well below the calibrated center-of-mass baseline (\textbf{2.91~px}) and well below chance (predict-center, \textbf{9.10~px}). Over the course of a clip the regression readout's error is below that of the COM baseline except for a brief transient about 250~ms after motion onset; from $\sim$570~ms into the clip it stays below the baseline for the rest of the motion (1.2~px versus 3.2~px averaged over 700--1850~ms; Fig.~\ref{fig:track}d). The calibrated center-of-mass is a \emph{linear} readout of the population centroid. That a learned spiking readout halves its error shows that the spiking-based computation accesses position information that the classical readout does not capture. Image reconstruction and object tracking are achieved by the same encoder module from the same spiking RGC population data; the tracking capability is thus a property of one system rather than of two separate SNN models.

\subsection{Energy efficiency}
From measured synaptic operations, our spike-driven decoder performs $\sim$$5.0\times10^{9}$ SynOps per sample (based on the moving-digit reconstruction model), versus $\sim$$6.8\times10^{10}$ dense MACs for the ANN-equivalent, giving \textbf{$\sim$70$\times$ lower compute energy at a 14.2\% mean firing rate} (Fig.~\ref{fig:efficiency}b). As detailed in Methods (Sec.~\ref{sec:energy}), this reduction in energy cost is the product of two terms: $\sim$13.6$\times$ reduction from spike sparsity and $\sim$5.1$\times$ reduction from the accumulate-versus-MAC cost convention (i.e.\ $13.6\times5.1\approx70$). This reduction applies to the actual computing cost only. The $\sim$5.1$\times$ cost term is fixed by the 45~nm CMOS semiconductor process node convention. The sparsity term, and hence the overall energy cost per sample, depends on the number of spike-driven synaptic operations, which the training objective can control. Exposing the firing rate as a differentiable penalty term (Sec.~\ref{sec:energy}) and training with $\lambda=0.1$ removes 21\% of the synaptic operations and raises the ratio to $\sim$88$\times$, at a cost in reconstruction fidelity (validation SSIM 0.80 $\to$ 0.73; Fig.~\ref{fig:efficiency}c). The penalty lowers firing where fan-out is largest, in the reconstruction decoder, and raises it in the small tracking head, so the mean firing rate is unchanged even as synaptic operations fall (Fig.~\ref{fig:efficiency}d). The spike count, not the mean rate, is therefore the relevant knob. Video~S4 shows this for one test clip, layer by layer: the spiking activity of each layer group in the $\lambda=0$ and $\lambda=0.1$ models, and the compute energy accumulating over the clip.

\subsection{Ablations}\label{sec:ablres}
We build the model up one component at a time, holding the shared-component ablation last (Table~\ref{tab:abl}; Fig.~\ref{fig:track}e and Fig.~\ref{fig:dynamic}f). Replacing the bottleneck
fully connected (F\_FC) tracking readout with the spiking centroid head
reduces tracking error from 4.1 to 2.6~px, and adding LIF temporal integration reduces frame-to-frame
jitter by roughly $10\times$ and improves accuracy to $\sim$2.2~px. For reconstruction, a channel--spatial--temporal
attention front-end adds $\sim$+0.04 SSIM, and the temporal-fusion readout (fusing all
per-timestep membranes rather than the last) raises dynamic reconstruction to
$\sim$0.80. The ablation is cumulative, and a dash marks a component that leaves the given metric unchanged
from the row above rather than an unmeasured value. The full model with the full-resolution regression
readout reaches reconstruction SSIM $\sim$0.80 and tracking error 1.33~px. Both numbers are measured from
the same trained model: the regression readout refines position without altering the reconstruction.

\begin{table}[htbp]\centering\small
\caption{Component ablation on dynamic-MNIST decoding. Rows are separate training runs.}\label{tab:abl}
\begin{tabular}{lcc}
\toprule
Configuration (cumulative) & Reconstruction SSIM & Tracking (px) \\
\midrule
F\_FC tracking readout & --- & 4.1 \\
Spiking centroid head & $\sim$0.69 & 2.6 \\
\;+ temporal integration (de-jitter) & --- & 2.2 \\
\;+ channel/spatial/temporal attention & +0.04 & 2.2 \\
\;+ temporal-fusion readout & $\sim$0.80 & 2.2 \\
\;+ full-resolution regression readout (full model) & --- & \textbf{1.33} \\
Task coupling (register / condition) & null (no gain) & 2.2--2.5 \\
\bottomrule
\end{tabular}
\end{table}

\section{Discussion}
The main goal in our work was to develop a neural decoding model, entirely in the spiking domain, especially for the purpose of reconstructing images of \emph{dynamical} visual targets, recorded ex vivo from a sizable population of mouse retinal ganglion cells (up to 1836 cells), and to do so computationally efficiently. An SNN model was built in which a single spike-native encoder (one architecture, trained separately for each stimulus set) reconstructs dynamic targets (flashed images, moving bars and drifting gratings, and randomly moving targets). In the case of spatially stationary, flashed images we show high-fidelity reconstruction across a range of natural and synthetic images.

The case of reconstruction and tracking of randomly moving targets (here simple handwritten digits moving in the $x$--$y$ plane) represents a particular challenge to any neural decoder. In our one-module SNN model the reconstruction is temporally refined by computational integration while the model simultaneously tracks the target movement. This is accomplished at a fraction of the ANN-equivalent computational energy.

The two experimental target scenarios, flashed stationary targets and randomly moving targets, stress the SNN model in different ways. The flashed stimuli are spatially rich, comprising natural scenes and synthetic patterns reconstructed at up to 512~px resolution. The moving targets in this paper are spatially simpler $32\times32$ digits; however, because their position is unknown and changes randomly, the SNN must integrate evidence over time to recover them. Reconstructing a target \emph{as it moves}, while sharpening the image as evidence accrues, is something a spike-native, temporally integrating network can do automatically, whereas a single-frame decoder fundamentally cannot. By contrast, recovering fine spatial structure from a briefly flashed, spatially constant natural scene poses a different problem, one that single-frame decoders can also meet, albeit at higher computational energy as shown above.

Our SNN decoder offers three advantages for a future visual neural interface. First, the same decoder architecture is able to \emph{reconstruct} moving as well as spatially stationary flashed stimuli recorded from large retinal spiking neural populations at high fidelity (masked-SSIM $\sim$0.81 natural, $\sim$0.88 synthetic), and does so as a compact SNN ($\sim$5.4M parameters). Second, the same encoder is capable of \emph{tracking} a moving target, approximately halving the error of a calibrated center-of-mass baseline model. Third, the decoder is computationally \emph{efficient}: it uses some 70-fold less compute energy than an ANN-equivalent, with spiking activity as a trainable `knob'. These capabilities can be understood by analogy: the decoder is effectively a \emph{biomimetic} computer that mimics the operating principles of the biological retina, using sparse, asynchronous spikes as the visual pathway does from the retina to image formation and perception in the brain.

The efficiency and event-driven structure of the decoder in this paper could suggest how a practical visual neural interface might be built in hardware. Because of the spike sparsity and addition-only synaptic operations, the network should be able to map efficiently onto asynchronous neuromorphic processors such as Intel Loihi and SynSense Speck-class event-driven systems-on-chip~\citep{davies2018loihi,yao2024speck} and other emerging silicon hardware whose energy scales with spiking activity rather than being locked to a global clock. On such hardware the roughly 14-fold estimated reduction in synaptic operations would translate into a comparable reduction in switching activity. The always-on, low-latency operation these devices provide also matches the demands of real-time, mobile, wearable visual prostheses where the compute-energy budget, not necessarily an algorithm's accuracy, can be the decisive consideration. Admittedly, we have focused on only one (critical) piece of a complete on-chip RGC prosthesis pipeline; mapping the full model to a specific device and measuring its wall-plug power and latency is a substantial engineering effort in its own right.

The `energy advantage' in this paper includes several elements. At a 14.2\% mean firing rate the decoder uses $\sim$70$\times$ less compute energy than the ANN-equivalent, as shown in Fig.~\ref{fig:efficiency}b, but this factor is \emph{activity dependent}. The overall efficiency can be further factored into two parts of different robustness: a $\sim$14$\times$ reduction in operation count from spike sparsity, which is independent of the silicon hardware process node, and a $\sim$5$\times$ reduction in per-operation cost (accumulate versus multiply--accumulate) under a 45~nm CMOS process convention. The $\sim$14$\times$ sparsity term is thus the conservative, node-independent core of the claim. As noted, the calculations in this paper are for compute-only costs. On real hardware, processes such as memory transfer, which per-operation accounting shows can dominate the energy budget~\citep{horowitz2014energy}, will reduce the realized energy advantage, and our estimates do not speak to full system-level power consumption.

This work has limitations. First, the results are bounded by the multielectrode
ex vivo recordings; improving the image reconstruction fidelity would require richer or multi-session data
rather than a larger decoder model. The dynamic stimuli in the form of moving handwritten
digits are relatively simple, and how our model handles \emph{complex natural scenes containing multiple
moving objects} remains an open question and a direction for future work. The computing energy estimates in this paper do not sweep the full accuracy--energy Pareto front. We have also evaluated richer spiking neuron
dynamics (adaptive-threshold neurons with spike-frequency adaptation), which appear not to improve
reconstruction over non-adaptive integrate-and-fire units. This justifies the use of the simpler LIF/IF neurons
and is consistent with a
particular neuron model not being the bottleneck. Future work could pursue several directions: adding confidence-gated early exit, so that
computation halts within a clip once the reconstruction and track have converged and fixed per-clip
compute becomes adaptive per-sample compute; extending to multi-trial and multi-session ex vivo or in vivo
recordings for calibrated uncertainty; and mapping the model onto emerging neuromorphic hardware to
measure on-chip power and latency.

\section{Methods}

\subsection{Benchtop retinal electrophysiology and data preparation}
Microelectrode array (MEA) recordings were conducted with the MaxOne HD-MEA system (Maxwell Biosystems, Switzerland). The HD-MEA is a CMOS-based recording platform comprising 26400 electrodes with an electrode pitch of 17.5~\textmu{}m. Up to 1024 electrodes can be simultaneously selected for extracellular recording, enabling high-resolution mapping of neuronal activity across large cell populations~\citep{ballini2014cmos}. Adult C57BL/6J mice were anesthetized with isoflurane (1–2\% at 0.5–1.0 L·min\textsuperscript{-1}). Enucleation was performed on one eye, which was quickly placed in oxygenated Ringer’s solution (in mM: NaCl 124, KCl 2.5, CaCl\textsubscript{2} 2, MgCl\textsubscript{2} 2, NaH\textsubscript{2}PO\textsubscript{4} 1.25, NaHCO\textsubscript{3} 26, and glucose 22; pH 7.35; oxygenated with 95\% O\textsubscript{2} and 5\% CO\textsubscript{2}). A small piece of retina (2 mm × 2 mm in size and 200~\textmu{}m in thickness) was excised and placed on the MEA, with the inner limiting membrane in contact with the MEA, to record the activity of retinal ganglion cells. A chamber surrounding the MEA was perfused with oxygenated Ringer’s solution at a rate of 1 mL·min\textsuperscript{-1}. Stimuli were delivered using a 405 nm DLP projector (PRO4500, Wintech, China). A stereomicroscope (SZ61, OLYMPUS, Japan) was used to focus the stimuli onto a 2 mm × 2 mm region of the photoreceptor layer of the retina. A total of 4000 distinct images were presented across 5 trials, with each image displayed for 400 ms, followed by 400 ms of darkness. Data were acquired at a sampling rate of 20 kHz. Raw data were initially sorted into different units using SpyKING CIRCUS~\citep{yger2018spyking} and then manually reviewed. High-quality single units were scored and selected based on mean pairwise correlation coefficient across trials (threshold: 0.2). Finally, the units were averaged across the 5 trials. Receptive-field mapping was performed by dividing the stimulus area (2 mm × 2 mm) into a 20 × 20 grid of small squares, which were sequentially illuminated in a randomized order. Each square was presented for 500 ms, followed by 500 ms of darkness. Spatial response heatmaps were generated separately for the ON window (during the 500-ms presentation) and the OFF window (during the subsequent 500-ms dark period) to determine the receptive field and ON/OFF response type.

\subsection{Retinal recordings and datasets}
All decoder inputs are population spike trains of mouse RGCs recorded on multielectrode arrays. The
\emph{flashed-image} regime comprises 1836 recorded RGCs responding to 4000 flashed images
(2000 synthetic and 2000 natural), binned at 10~ms over an 80-bin ($0.8$~s) response window
(400~ms of image presentation followed by 400~ms of blank), with six repeat trials per image. The \emph{dynamic} regime comprises 1206 RGCs responding to 3000 clips of
randomly moving handwritten digits, binned at 10~ms over a 200-bin (2~s) window; each clip presents a
single digit executing an independent, \emph{completely random} $x$--$y$ trajectory (1.6~s of motion
followed by 0.4~s of blank), so the target moves unpredictably across the field. From these spikes the
network jointly decodes two things: it \emph{tracks} the digit's time-varying position and, from the same
shared encoder, \emph{reconstructs} the target. We
also decode other dynamic stimuli: eight-direction moving bars and 40 drifting
gratings (a separate recording of 760 RGCs). Both are decoded from a causal sliding window over the recording, 10 bins (100~ms) long for the bars and 3 bins (30~ms) for the gratings. Each of these recordings comprises ten repeat trials; for the bars, five trials were used for training, one for validation and four were held out for testing, and the reported bar metrics are computed over windows in which the bar lies inside the recorded $32\times32$ patch. Held-out evaluation uses a fixed
random split, and every reported number is
computed on data disjoint from training.

\subsection{From spikes to a spatial input tensor}
A recorded population is an unordered set of cells at irregular positions on the retina, whereas a
convolutional network requires its input on a regular spatial grid. How that mismatch is bridged
determines what spatial structure the network is able to exploit at
all: an arbitrary ordering of cells into channels would discard the retinotopic organization that the
population already carries.

Because RGCs tile the retina, each cell's spikes report activity at a specific visual-field location. We
therefore place each cell's firing rate at its measured receptive-field center and spread it with an
isotropic two-dimensional Gaussian kernel of standard deviation $0.5$ grid pixels (\emph{Gaussian splatting}~\citep{westover1990footprint}). Each kernel is normalized to unit mass
per neuron, so every cell contributes equally regardless of where its center falls relative to the pixel
lattice (Fig.~\ref{fig:concept}c). Where receptive fields overlap, contributions are summed rather than
averaged, so that neighboring cells reinforce one another and sub-pixel differences in receptive-field
position are preserved instead of being collapsed onto the nearest grid point (a hard snap; compare the two panels of Fig.~\ref{fig:mea}d). This yields an input tensor
of shape $[\,T\times C\times 32\times 32\,]$ where $T$ is the
number of time bins and $C$ is the number of polarity channels. Describing retinal receptive fields as
two-dimensional Gaussians is long established in retinal physiology~\citep{rodieck1965quantitative,chichilnisky2001white}; here that description is used in the
opposite direction, as the rasterization step that turns an irregularly sampled population into a decoder
input. Cells are separated into ON, OFF, and
ON--OFF channels ($C=3$) for the dynamic-digit recording, reflecting the retina's parallel encoding of
contrast polarity~\citep{wassle2004parallel,krieger2017alpha}. Polarity labels were not available for the
flashed-image recording, which was decoded without polarity separation. This
receptive-field-mapped, polarity-separated representation preserves the retinotopic and functional
structure of the population while presenting the network with a spatially patterned ``spiking movie''.

\subsection{Spiking Neural Network}

\subsubsection{Shared SNN encoder}
The core of the SNN is a spiking convolutional U-Net-like structure~\citep{ronneberger2015unet} (Fig.~\ref{fig:arch}) built from
leaky or non-leaky integrate-and-fire (LIF/IF) neurons, implemented in
SpikingJelly~\citep{fang2023spikingjelly} with a CUDA/cupy backend for training and evaluation. Each unit
maintains a membrane potential $V$ that is updated at every time step $t$ as
\begin{align}
H[t] &= \beta\, V[t-1] + X[t], \\
S[t] &= \Theta\!\left(H[t] - V_{\mathrm{th}}\right), \\
V[t] &= \left(1 - S[t]\right) H[t],
\end{align}
where $X[t]$ is the input current at step $t$, given by the convolution of the preceding layer's spikes with
the layer weights; $\Theta$ is the Heaviside step function; $S[t]\in\{0,1\}$ is the emitted spike;
$V_{\mathrm{th}}=1$ is the firing threshold; and $\beta$ sets the membrane leak, with $\beta = 1$ for a
non-leaky IF unit and $\beta = 1 - 1/\tau$ for an LIF unit with membrane time constant $\tau$. A spike
resets the potential to zero. Because $\Theta$ has zero
derivative almost everywhere, the backward pass uses a surrogate gradient~\citep{neftci2019surrogate}.

The encoder downsamples three times ($32\!\to\!16\!\to\!8\!\to\!4$) with feature widths of
$48\!\to\!96\!\to\!192\!\to\!384$ channels and optional residual connections. Each stage widens the region
of visual space to which a unit responds, so the deepest features summarize the field as a whole while the
shallowest retain the retinotopic detail of the input. This division of labor matters because the input is
a coarse and noisy $32\times32$ rate map, not a photographic image: the $4\times4$ bottleneck alone
cannot support stroke-level reconstruction, and the skip connections that carry shallow features forward to
the reconstruction module return that detail to the output.

In depth the network is a feedforward feature hierarchy; in time it
is a stateful integrator, because every unit carries its membrane potential from one time bin to the next.
The same weights are applied at each of the $T$ steps, so across a clip the computation is fixed and only
the accumulated state changes. Evidence about the stimulus therefore accumulates as the clip proceeds, and for a
moving target the reconstruction can sharpen without any explicit re-centering of the target on each frame.
This motion invariance comes from the network's temporal memory, not from a preprocessing step.

Each unit emits a binary value,
so graded stimulus intensity cannot be carried by activation amplitude as it would be in an artificial
network; it must instead be represented by which units fire and at which time steps, distributed across the
population and across the $T$ bins. This is why the reconstruction is read from an accumulated membrane
potential rather than from a spike count. In exchange, computation stays event-driven and sparse
(Sec.~\ref{sec:energy}), which suits the model to event-driven neuromorphic hardware and keeps the
learned mapping in the same units the retina itself uses. Apart from the input stage, which reads the
RF-mapped input (the optional attention front-end and the first convolution), every layer receives spikes
(binary or, where a residual block adds its two spike branches, counts of at most two), so its synaptic
operations are accumulates. The remaining analog operations decode the membranes of the non-firing
readout neurons.

\subsubsection{SNN reconstruction module}
The SNN reconstruction module mirrors the encoder with three nearest-neighbor upsampling stages, which keep the
upsampled tensor a spike tensor, and skip connections that reintroduce
fine spatial detail. Its output layer is a single high-threshold (5000) analog-membrane readout neuron
whose accumulated membrane potential, passed through a sigmoid, is the reconstructed image
$[\,1\times H\times H\,]$ at the target resolution $H$ (32 for the dynamic experiments; up to 512 for the
flashed natural/synthetic experiments). Training minimizes a combined loss~\citep{qin2019basnet}
$\mathcal{L} = \mathcal{L}_{\mathrm{BCE}} + \mathcal{L}_{\mathrm{SSIM}} + \mathcal{L}_{\mathrm{IoU}}$
(equal weights), where $\mathcal{L}_{\mathrm{SSIM}}$ uses a window size of 3 (matching the small target
structures) and $\mathcal{L}_{\mathrm{IoU}}$ encourages correct support. Optimization uses AdamW~\citep{loshchilov2019adamw} at a
base learning rate of $10^{-3}$ with weight decay $10^{-4}$ and a reduce-on-plateau schedule; mixed-precision
(fp16) autocast is used throughout. The same end-to-end spiking model (trained separately for each dataset) is used for the flashed
natural/synthetic and the dynamic reconstructions. The high-resolution flashed-image results are obtained by
training it at the larger target resolution on the full 1836-cell population, with the four additional
$\times2$ upsampling stages this requires interleaved with spiking convolutions.

\subsubsection{SNN tracking module}
The moving target's position is read out by the SNN tracking module, a lightweight spiking branch attached
to the encoder's finest ($32\times32$) feature map $F$. The branch applies a spiking convolution, batch
normalization and an LIF/IF layer, followed by a $1\times1$ convolution that collapses the result to a
single channel over the $32\times32$ positions and drives a layer of LIF neurons, whose spikes form the
activity map $M[t]$. Because per-timestep maps are noisy, the LIF membrane integrates its input over the
preceding bins ($\tau = 6$ bins) before a unit fires; this
suppresses frame-to-frame jitter by roughly an order of magnitude.

Position is then obtained as the spike-count (spike-weighted) centroid of the map,
\begin{equation}
p_i[t] = \frac{c_i[t]}{\sum_j c_j[t]}, \qquad c_i[t] = \gamma\, c_i[t-1] + M_i[t],
\qquad
(x,y)[t] = \sum_i p_i[t]\, \mathbf{r}_i ,
\end{equation}
where $\mathbf{r}_i$ is the normalized coordinate of position $i$ on the grid and $c_i[t]$ the leaky spike
count of unit $i$ (decay $\gamma$). Both sums are accumulated by non-firing readout neurons with fixed
synaptic weights, so only the final ratio is analog. This
is a learned, spiking population-vector (center-of-mass) readout~\citep{georgopoulos1986population}.

The centroid reduces the whole feature map to a single point and therefore discards any structure that
is not captured by it. A second, more expressive readout recovers that structure by predicting an offset
from the centroid,
\begin{equation}
(x,y)[t] = \underbrace{\textstyle\sum_i p_i[t]\,\mathbf{r}_i}_{\text{centroid}}
\;+\; s\,\tanh\!\big(g(F[t])\big),
\end{equation}
where $g$ is a small spiking network (LIF integration, two stride-2 spiking convolutions and a spiking
fully connected layer) applied to the \emph{entire} finest feature map, whose output drives two non-firing
readout neurons, and $s = 0.5$ bounds the
correction. The final layer of $g$ is initialized to zero, so at the start of training the readout reproduces
the centroid exactly and thereafter learns only a correction to it. In this sense the centroid
acts as a \emph{coarse prior}: it supplies a reliable first estimate, and the regression refines it. This
full-resolution regression readout is our best tracker.

Both readouts are trained by minimizing the mean squared error between predicted and true normalized
coordinates, averaged over the valid part of each clip. The target is the true stimulus trajectory rather
than any property of the neural response, shifted forward by 25 time bins ($250$~ms) to allow for retinal
response latency. The loss is masked to the bins over which the shifted trajectory is defined.
Performance is reported as the mean Euclidean position error in $32$-grid pixels.

As a non-learned reference we compute the classical center-of-mass (COM) of the population
activity accumulated over the preceding 15 bins (150~ms) and calibrate it to true position with a least-squares
affine map (a $2\times2$ gain matrix plus offset), fit on 200 clips of the recording. We refer to this as the \emph{calibrated COM baseline}. It is a strong, standard readout, not an
information-theoretic optimum.

\subsubsection{Task coupling (ablation only)}
We optionally couple the two modules so that the tracked position modulates the reconstruction
features, either by a soft Gaussian gate centered on the estimated position (\emph{condition}) or by a
grid-sample warp that re-registers features toward the center (\emph{register}). The coupling is
\emph{bidirectional} for \emph{condition}: the reconstruction loss back-propagates through the gate into the tracking
head, so in principle the two tasks train one another. For \emph{register}, the position estimate is detached from the
warp grid, so the tracker still learns true position while the warp uses those coordinates. Coupling
is evaluated purely as an ablation (Sec.~\ref{sec:ablres}) and is not part of the primary model; as reported there, it
does not improve reconstruction, and we include it for completeness rather than as a contribution.

\subsubsection{Attention front-end and temporal-fusion readout}
Two further components sharpen reconstruction, and both are ablated in Sec.~\ref{sec:ablres}. First, a
\emph{multi-dimensional attention} front-end (MA-SNN~\citep{yao2023masnn}) re-weights activity in the input spike tensor
before the encoder, jointly along the channel (ON/OFF/ON--OFF polarity), spatial, and
temporal axes: the three axes along which retinal information is organized. This lets the otherwise purely
local convolutional encoder exploit non-local spatial, cross-polarity, and temporal structure it would
miss. Second, rather than
reading the reconstruction from the output membrane at the final timestep alone, a \emph{temporal-fusion
readout} learns a softmax weighting over the per-timestep output membranes and reconstructs from their
weighted sum. It is initialized, through a positional bias, to reproduce the last-timestep readout exactly,
and then learns to redistribute weight toward the most informative moments of the clip. Because a
randomly moving target is best resolved at different times during its trajectory, fusing evidence across
time rather than sampling a single frame is the component that carries dynamic reconstruction past the
last-timestep baseline (Sec.~\ref{sec:ablres}).

\subsection{Firing-rate objective and energy estimation}\label{sec:energy}
Because energy on neuromorphic hardware scales with the number of spikes, we estimate compute energy from synaptic
operations (SynOps) counted over a forward pass using SpikingJelly's operation counter, under a 45~nm
CMOS convention in which a multiply--accumulate (MAC) costs $4.6$~pJ and an accumulate (AC) costs
$0.9$~pJ~\citep{horowitz2014energy}.

A synaptic operation is a single accumulation of a weight into a postsynaptic potential, triggered by a
presynaptic spike. For a layer $\ell$ in which each presynaptic unit drives $f_\ell$ synapses,
\begin{equation}
\mathrm{SynOps} \;=\; \sum_{\ell} f_\ell \sum_{t=1}^{T} \big\lVert S_\ell[t] \big\rVert_0 ,
\end{equation}
where $\lVert S_\ell[t]\rVert_0$ is the number of units of layer $\ell$ that spike at step $t$. The work
therefore scales with the number of spikes rather than with the size of the layer: a unit that stays silent
costs nothing. In a dense network the same operation is a multiply--accumulate, because both operands are
real-valued. When the presynaptic value is a binary spike, the multiplication degenerates to a copy and only
an accumulate is required; this is the origin of the per-operation saving. SynOps counts only the
accumulations that arise from synaptic transmission, whereas the broader accumulate count also includes
residual additions, normalization and pooling. We report the SynOps-based estimate. The spike-driven SNN performs $\sim$$5.0\times10^{9}$ SynOps per
sample, versus $\sim$$6.8\times10^{10}$ dense MACs for an ANN of identical architecture (the
``ANN-equivalent''): the same compact $\sim$5.4M-parameter network evaluated with dense, clocked arithmetic.
This is an \emph{operational} energy comparison (synaptic operations versus dense multiply--accumulates for
one fixed architecture and parameter count), not a claim about a separately trained ANN's reconstruction
fidelity. The resulting energy ratio of $\sim$$70\times$ decomposes into a
$\sim$$13.6\times$ reduction in operation count (spike sparsity) and a $\sim$$5.1\times$ reduction in
per-operation cost (accumulate versus multiply--accumulate). We report this decomposition explicitly, and
note that the estimate is compute-only and excludes memory-movement energy.

Firing rate is additionally exposed as a differentiable penalty term, so that sparsity can be optimized
rather than merely measured. During each forward pass we record the mean activation of every spiking node
and average it over nodes and timesteps,
\begin{equation}
\bar{r} \;=\; \frac{1}{NT}\sum_{n=1}^{N}\sum_{t=1}^{T} \overline{s_{n}(t)},
\end{equation}
where $\overline{s_{n}(t)}$ is the fraction of units firing in spiking layer $n$ at timestep $t$ and $N$ is
the number of spiking layers. This quantity retains its gradient and is added to the training objective,
\begin{equation}
\mathcal{L} \;=\; \mathcal{L}_{\mathrm{recon}}
\;+\; \lambda_{\mathrm{track}}\,\mathcal{L}_{\mathrm{track}}
\;+\; \lambda\,\bar{r},
\end{equation}
so that raising $\lambda$ penalizes dense firing directly and moves the network along the
accuracy--energy trade-off. A single penalized model ($\lambda=0.1$) is reported in Fig.~\ref{fig:efficiency}c,d; a full swept characterization of that
trade-off, in which $\lambda$ is varied systematically, is future work.

\subsection{Metrics and evaluation protocol}
Reconstruction is quantified by the structural similarity index (SSIM), peak signal-to-noise ratio (PSNR)
and mean squared error (MSE); for the high-resolution flashed-image decodes we additionally report a
foreground-masked SSIM (``masked-SSIM,'' computed only over the object support to avoid rewarding the large
black background) and the LPIPS perceptual distance~\citep{zhang2018lpips} (a learned feature-space image-similarity metric; lower
is better). SSIM is reported with a Gaussian window of 11 for the 512-px flashed-image decodes, following
the standard convention~\citep{wang2004ssim}, and with a window of 3 for the $32$-px moving-stimulus
reconstructions (bars, gratings and randomly moving digits), where the target occupies only a few pixels. The training loss uses a window of 3
throughout. The flashed natural and synthetic numbers are reported as masked-SSIM, the foreground-only
convention appropriate to the large black background. Tracking is quantified by the mean Euclidean position
error in $32$-grid pixels over the valid response window. All evaluation is on held-out test clips.

\section{Conclusion}
We have presented an end-to-end spiking neural network that reconstructs visual stimuli directly from the
population activity of retinal ganglion cells, and that tracks a moving target from the same shared encoder.
The model reconstructs flashed natural and synthetic images at masked-SSIM $0.81$ and $0.88$ from 1836
cells, and moving stimuli at SSIM $0.93$ for bars, $0.80$ for drifting gratings and $\sim$$0.80$ for
randomly moving digits. The reconstruction is refined over time as the spiking membranes accumulate
evidence during motion. A learned spiking readout locates the moving target to $1.33$~px, halving the error
of a calibrated center-of-mass baseline. Reconstruction fidelity is comparable across 18 architectures spanning
more than two orders of magnitude in size, suggesting a data-limited rather than a capacity-limited regime. Our network
reaches that fidelity with $\sim$5.4~M parameters and, being sparse and event-driven, at a small fraction of
the compute energy of an equivalent dense network. Decoding dynamic vision in the spike domain therefore appears
feasible at a computational cost compatible with the implantable and wearable interfaces for which such decoding is
ultimately intended.

\section*{Supplementary material}
\noindent\textbf{Video S1. Moving bars and gratings.} Moving bar (real time) and drifting gratings (1--15~Hz, half speed) reconstructed from RGC spikes on test trials; the grating example is the best of 32 test cases.

\noindent\textbf{Video S2. Randomly moving digits.} Ten selected test clips of randomly moving handwritten digits in real time: stimulus, RGC input, and the reconstruction drawn at the position decoded by the same network.

\noindent\textbf{Video S3. Flashed images over time.} Six flashed test images from Fig.~\ref{fig:flashed} (three synthetic, three natural): the reconstruction builds up as spikes accumulate over one 800-ms trial, scored by pixel correlation.

\noindent\textbf{Video S4. Sparse spiking computation.} One test clip through the model trained without and with the firing-rate penalty ($\lambda=0.1$): per-layer spiking activity at each 10-ms step and the accumulating compute energy.

\section*{Declarations}
\textbf{Acknowledgment.} The authors are very grateful for the many insights contributed by Leo Kozachkov, and to Jiaxin Lei and Rebekah Zhao at Brown for their technical expertise. We owe a special debt to Weilong Li, Yiheng Wang, Qingyuan Chen and Jiayi Zhang for the superb experimental work and willingness to share the experimental data for this paper. The research at Brown was supported by the National Science Foundation Collaborative Award \#232600. We also acknowledge the computational resources supported by the NIH S10 instrumentation grant S10OD036341 (High-Performance Compute Cluster for Brain Science) via the Center for Computation and Visualization (CCV) at Brown University.
\\\\
\textbf{Data availability.} Preprocessed RGC spike tensors and test split indices are available from the authors on reasonable request; raw recordings are subject to institutional data-sharing agreements. Model, training, evaluation, and energy-measurement code are available at
\url{https://github.com/nzhang1218/snn_rgc_decoding.git}.
\\\\
\textbf{Conflict of interest.} The authors declare no conflict of interest.

\bibliographystyle{unsrtnat}
\bibliography{references}
\end{document}